\documentclass[pdflatex,sn-mathphys-num]{sn-jnl_adjusted}

\usepackage{graphicx}%
\usepackage{multirow}%
\usepackage{amsmath,amssymb,amsfonts}%
\usepackage{amsthm}%
\usepackage{mathrsfs}%
\usepackage[title]{appendix}%
\usepackage{xcolor}%
\usepackage{textcomp}%
\usepackage{manyfoot}%
\usepackage{booktabs}%
\usepackage{algorithm}%
\usepackage{algorithmicx}%
\usepackage{algpseudocode}%
\usepackage{listings}%
\usepackage{siunitx}%
\begin{document}
	\title[Article Title]{Terahertz Quantum Imaging}
	
	
	\author*[1,2]{\fnm{Mirco} \sur{Kutas}}\email{mirco.kutas@itwm.fraunhofer.de}
	
	\author[1]{\fnm{Felix} \sur{Riexinger}}
	
	\author[1]{\fnm{Jens} \sur{Klier}}
	
	\author[1]{\fnm{Daniel} \sur{Molter}}
	
	\author*[1,2]{\fnm{Georg} \sur{von Freymann}}\email{georg.von.freymann@itwm.fraunhofer.de}
	
	\affil[1]{\orgdiv{Department Materials Characterization and Testing}, \orgname{Fraunhofer Institute for Industrial Mathematics ITWM}, \orgaddress{\city{Kaiserslautern}, \country{Germany}}}
	
	\affil[2]{\orgdiv{Department of Physics and Research Center OPTIMAS}, \orgname{RPTU Kaiserslautern-Landau}, \orgaddress{\city{Kaiserslautern}, \country{Germany}}}
	
	
	\abstract{Quantum imaging with undetected photons spatially transfers amplitude and phase information from one spectral region of physical interest to another spectral region that is easy to detect. The photon energy of the two spectral regions can, in principle, be separated by several orders of magnitude. However, quantum imaging with undetected photons has so far only been demonstrated in spectral regions of similar order of magnitude in frequency (and for which cameras are commercially available). Here, we demonstrate amplitude- and phase-sensitive imaging in the terahertz spectral region (\SI{1.5}{\tera\hertz} center frequency) by detecting only visible photons (center wavelength \SI{662.2}{\nano\meter}, \SI{452.7}{\tera\hertz} center frequency) more than two orders of magnitude apart. As a result, terahertz spectral information can be reliably detected with a standard CMOS camera without cooling, achieving a spatial resolution close to the wavelength. By taking advantage of quantum distillation in a nonlinear interferometer, the influence of ubiquitous thermal terahertz photons can be neglected. Our results are in good agreement with numerical simulations of the imaging process and demonstrate the huge potential of this method to address otherwise challenging spectral regions where cameras do not exist.}
	

	\maketitle
	
	\newpage
	
	\section{Introduction}\label{introduction}
	
	Quantum imaging is one of the most important pillars of quantum technology and is driven by the promise to surpass the limitations of classical imaging techniques. To date, realizations range from super-resolution imaging \cite{Classen.2017, Unternahrer.2018,Tenne.2019}  over imaging with a small quantity of photons \cite{Aspden.2013, Morris.2015, Gregory.2020} to noise-reduced imaging \cite{Brida.2010, SabinesChesterking.2019, Samantaray.2017}. Furthermore, established classical imaging methods have been advanced based on the unique properties of quantum light towards innovative imaging solutions such as distillation \cite{Defienne.2019,Fuenzalida.2023} and holography \cite{Defienne.2021,Topfer.2022,Haase.2023}. One of the most promising implementations is quantum imaging with undetected photons \cite{Lemos.2014}, as it enables the decoupling of the spectral range of measurement and the spectral range of detection. Therefore, non-degenerate photon pairs (biphotons) are generated via spontaneous parametric down-conversion (SPDC), with the measured photon referred to as the signal and the one discarded as the idler.
	
	This measurement principle was initially demonstrated based on the effect of induced coherence without induced emission \cite{Wang.1991, Zou.1991} and generally relies on the indistinguishability of optical paths in a nonlinear interferometer \cite{Hochrainer.2022}. In a single-crystal layout (Michelson-geometry), this means that distinguishing between biphotons generated during the first and second pass of the pump radiation through the nonlinear medium becomes impossible after the second pass. As a result, interference occurs in the individual spectral components, which mutually influence each other \cite{Chekhova.2016}. This phenomenon allows for the visualization of losses or phase changes experienced by the idler photons between the passes through the crystal, even though only the associated signal photons are measured. The direct access to amplitude and phase information is enabled by the interferometric nature of this technique. Numerous experiments have already demonstrated the high potential of this technique for various tasks, particularly in the infrared spectral range \cite{Kviatkovsky.2020, Paterova.2020, GilaberteBasset.2021, Paterova.2021}. To this date, the largest spread between signal and idler in the context of imaging has been achieved with idler photons at the lower limit of the mid-infrared spectral range \cite{Kviatkovsky.2022}.
	
	However, the true potential of this method is revealed in spectral ranges where multi-pixel detectors are not readily accessible, such as the terahertz frequency range. Here, direct imaging techniques are technically challenging, as they either involve heavily cooled detectors \cite{Simoens.2014} or the implementation of compressed sensing schemes \cite{Chan.2008}. Despite these challenges, terahertz imaging plays a major role in industry and science, as it provides direct access to the amplitude and phase information of the electric field and can penetrate a wide range of substances. These capabilities open a variety of applications in security inspection \cite{Federici.2005}, quality control \cite{Ellrich.2020}, biomedical imaging \cite{Pickwell.2006}, and art conservation \cite{Fukunaga.2008}. To avoid direct detection, the transfer of intensity images from the terahertz to the visible spectral range was investigated \cite{Fan.2015, Downes.2020}. However, these approaches lack access to the phase information of the terahertz wave. On the other hand, coherent detection schemes enabling this access \cite{Jiang.1999} have never been established despite decades of development. The development of quantum-based measurement methods, however, has only begun.
	
	In this work, we present quantum imaging with undetected photons in the terahertz frequency range using components for generation and detection that work exclusively in the visible range. The image is acquired by solely detecting the signal photons with a camera using a quantum distillation scheme and provides access to the amplitude and phase of the terahertz radiation. This enables us to obtain images in the terahertz frequency range with a resolution of up to \SI{240}{\micro\meter} close to the used wavelength of $\lambda_\mathrm{THz} = \SI{200}{\micro\meter}$. To further evaluate the imaging parameters of quantum imaging in the terahertz frequency range, we theoretically simulate the setup of a single-crystal quantum interferometer with idler radiation in the terahertz frequency range. 
	
	\section{Results}\label{results}
	
\begin{figure}
		\centering
		\makebox[\textwidth][c]{\includegraphics[width=\linewidth]{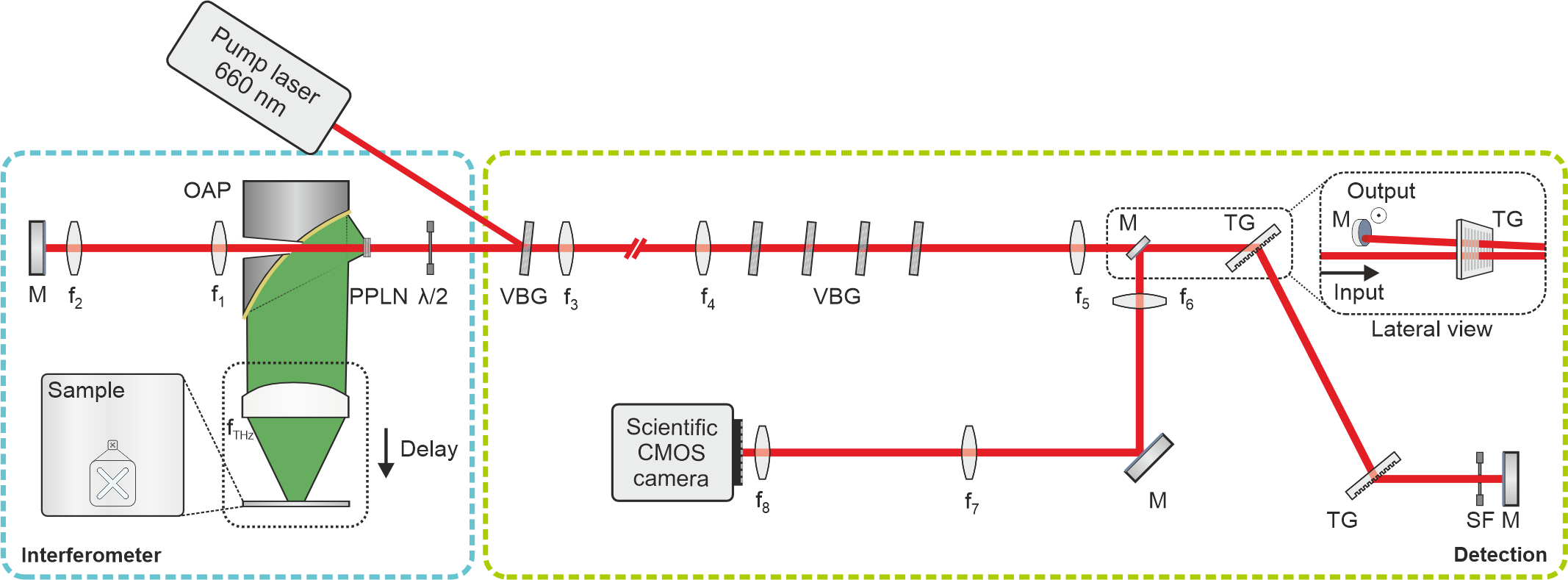}}
		\caption{Schematic of the experimental setup. A continuous-wave laser at \SI{660}{\nano\meter} is reflected by a volume Bragg grating (VBG) into the interferometer part of the setup, generating highly non-degenerate biphoton pairs in a periodically poled 1-mm-long MgO:LiNbO$_3$ crystal. The pairs are separated by an off-axis parabolic mirror (OAP) with a through hole. At this mirror, the terahertz radiation is reflected while signal and pump radiation are transmitted to ensure the sample is exclusively illuminated with the terahertz radiation. The path length differences (delay) of the nonlinear interferometer can be varied via a movable linear stage on which the sample and the preceding lens (f$_\mathrm{THz}$) are mounted. After the second traverse, the signal is imaged via several 4f lens systems onto the sCMOS camera. To solely image the signal radiation, the pump radiation is filtered by several VBGs. Additionally, generated parasitic signal radiation is spatially distributed by applying two transmission gratings (TG) in series and suppressed via a spatial filter (SF). The filter scheme's side view (inset) indicates that neither dichroic optics nor a beam splitter are introduced into the beam path. The length of the optical paths is not to scale.}
		\label{fig:setup}
	\end{figure}

	The core of our experimental setup, shown in Fig.~\ref{fig:setup}, is formed by a Michelson-like single-crystal nonlinear interferometer in position correlation configuration. In contrast to a nonlinear interferometer exploiting momentum correlation, as initially demonstrated by Lemos \textit{et al}. \cite{Lemos.2014}, a setup utilizing position correlation enables a significantly enhanced optical resolution \cite{Kviatkovsky.2022, Viswanathan.2021, GilaberteBasset.2023}. To generate biphoton pairs with one photon in the terahertz frequency range, a \SI{1}{\milli\meter}-long periodically poled magnesium oxide doped lithium niobate (MgO:LiNbO$_3$) crystal is illuminated by a frequency-doubled solid-state laser operating at \SI{660}{\nano\meter}. The poling period of \SI{72}{\micro\meter} results in the generation of terahertz photons with a center frequency of \SI{1.5}{\tera\hertz} and associated signal photons with a wavelength of approximately \SI{662.2}{\nano\meter}. The separation of the generated biphotons is realized by an off-axis parabolic mirror with a through hole placed in focal distance to the crystal, ensuring the sample is solely illuminated by terahertz radiation. In our setup, this is achieved purely geometrically, as the emission angles of the terahertz radiation from the nonlinear crystal exceed those of the signal \cite{Kutas.2020}. All generated photons, as well as the pump radiation, are first imaged onto the end mirror of the respective interferometer arm using a 4f lens system and then back into the image plane of the crystal. Methods provides further details on the alignment process of the nonlinear interferometer. After the second pass through the crystal, the biphotons are indistinguishably superimposed, allowing the terahertz radiation to be neglected from this point onwards, as the signal radiation carries the information about the changes the idler experienced. To image the signal onto the camera, several 4f lens systems are connected in series (see lenses $\mathrm{f}_3$ to $\mathrm{f}_8$ in Fig.~\ref{fig:setup}), enabling the image plane of the crystal to be imaged directly onto the camera plane.

	To observe the weak signal radiation with the camera, isolating it from other contributions is necessary. In addition to the pump radiation, these include parasitic signal components that are typical for the generation of terahertz radiation by parametric down-conversion \cite{Haase.2019, Kitaeva.2011} (Methods). Therefore, volume Bragg gratings (VBG) and a spectral filter scheme are implemented in the detection part. The used VBGs are spectrally narrow-band notch filters that suppress the pump radiation by reflection but transmit even slightly frequency-shifted radiation almost completely \cite{Haase.2019}. These filters are very sensitive to the radiation`s angle of incidence and are, therefore, positioned in parts of the optical system where the pump radiation is collimated. To isolate the signal radiation from parasitic contributions at different wavelengths, two transmission gratings are positioned in series, converting the spectral distribution of the radiation into a spatial distribution. A vertical slit in the beam path spatially --- and consequently spectrally --- filters the parasitic contributions. To separate the incident and the reflected (filtered) radiation, the reflecting mirror is slightly tilted, as illustrated in the lateral view in Fig.~\ref{fig:setup}.
	
	In contrast to other spectral regions, the signal interference visibility for quantum sensing experiments performed in the terahertz frequency range is very low (in our case below \SI{0.15}{\percent} per pixel) due to the unfavorable properties of the used nonlinear crystal \cite{Kutas.2020, Kutas.2021}. The large refractive index ($n_\mathrm{THz} = 5.1$) and the high absorption ($\alpha$ = \SI{30}{\per\centi\meter}) of MgO:LiNbO$_3$ in the terahertz frequency range \cite{Wu.2015} result in significant losses of the idler radiation within the nonlinear medium. Moreover, parasitic signal photons are also generated at the signal wavelength as a consequence of thermal terahertz radiation present at room temperature. Following \cite{Fuenzalida.2023}, this results in a signal-to-noise ratio of at most 1:700 in our experiment. Nevertheless, MgO:LiNbO$_3$ is a plausible choice as a nonlinear material for a proof-of-principle experiment, as it is well-known and widely used for the nonlinear generation of terahertz radiation due to its high nonlinear coefficient and comparatively easy handling.
	
	\begin{figure}
		\centering
		\makebox[\textwidth][c]{\includegraphics[width=\linewidth]{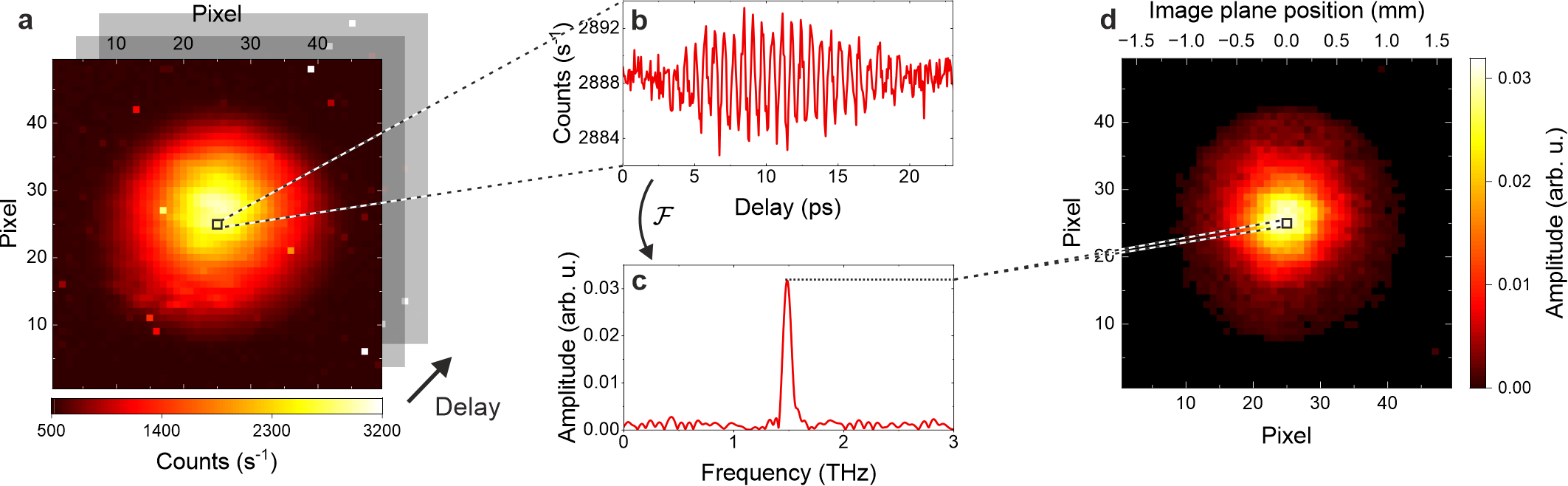}}
		\caption{Quantum distillation scheme. (a) Camera image of the signal with a 3$\times$3 pixel binning performed. The image represents an average of 1000 images with an exposure time of \SI{1000}{\milli\second} each at a fixed delay. Semi-transparent image layers display images that were captured with a larger delay. The used crystal provides a poling period of $\Lambda$ = \SI{72}{\micro\meter} leading to generated terahertz frequencies of about \SI{1.5}{\tera\hertz}. (b) Observed waveform and (c) corresponding fast Fourier transform of the pixel highlighted in (a). The quantum distilled image (d) is obtained by determining the maximum amplitude value of the FFT in the area of the expected interference (\SIrange{1.4}{1.6}{\tera\hertz}) and placing this value at the position of the corresponding pixel. A simple metal plate is used as a sample.}
		\label{fig:acquisition}
	\end{figure}
	
	To extract the terahertz quantum image from the dominant superposition of the intrinsic classical noise, we use a quantum distillation scheme based on the interference frequency, as illustrated in Fig.~\ref{fig:acquisition}. The signal radiation is recorded with an sCMOS camera that provides a quantum efficiency of \SI{55}{\percent} per pixel (\SI{5.4}{\micro\meter}$\times$\SI{5.4}{\micro\meter}) at \SI{660}{\nano\meter}. For the camera, a 3$\times$3 pixel binning is applied (see Fig.~\ref{fig:acquisition}a). The individual images are recorded with an exposure time of \SI{1000}{\milli\second}, a relative shift of \SI{10}{\micro\meter} (corresponding to a delay of about \SI{0.07}{\pico\second}), and averaged over 1000 images to further reduce noise contribution. By evaluating a single pixel over the entire delay, we obtain a waveform of the observable interference, as exemplified in Fig.~\ref{fig:acquisition}b. This waveform contains phase and amplitude information of the corresponding terahertz radiation. To extract this information, the waveform's fast Fourier transform (FFT) is calculated (see Fig \ref{fig:acquisition}c). Subsequently, the frequency spectrum is analyzed in a \SI{0.2}{\tera\hertz} range around the associated central frequency of \SI{1.5}{THz}. The maximum amplitude value within this range is assigned to the evaluated pixel in the final image. Fig.~\ref{fig:acquisition}d shows the result of the quantum distillation performed for a measurement taken with a simple metal plate, which will be used as the reference image in the following.
	
	To characterize our imaging system, we analyze the field of view (FoV) and the resolution. Following the definition in \cite{Kviatkovsky.2022}, where the FoV is defined as the full width at half maximum of the distribution illuminating the sample, we obtain $FoV_\mathrm{theo}=\SI{1.3}{\milli\meter}$. However, relating the images obtained to a reference image also permits extracting information about the sample that falls outside this definition. In this case, the FoV is limited by the noise level after performing the FFT. To reduce noise in areas with low counts, we only take pixels into account for which the amplitude in the reference image (see Fig.~\ref{fig:acquisition}d) is $1\times 10^{-3}$ above the noise value and obtain $FoV_\mathrm{exp}=\SI{2.0 \pm 0.2}{\milli\meter}$ (Methods).
	The difference between the theoretical and experimental FoVs shows the need for a more detailed description of the experiment. To achieve this, we use a numerical simulation method for quantum sensing setups \cite{Riexinger.2023}, adapted to position correlated imaging, to simulate the detector images. The transition amplitudes for the biphoton states are based on \cite{Viswanathan.2021b} and evaluated with a quasi-Monte Carlo integration method developed for the simulation of SPDC sources \cite{Riexinger.2023b}. Methods provides a detailed description of the model.
	Using the simulated detector images, we obtain the characterizing quantities for the setup following the same steps as with the experimental data. With the same cutoff level, we obtain $FoV_\mathrm{sim}=\SI{2.2 \pm 0.1}{\milli\meter}$, matching the experimental value much better.
	For resolution determination, a knife-edge measurement is carried out in the image plane of the terahertz beam path (Methods). According to the definition in \cite{Kviatkovsky.2022}, we determine the resolution of the system to $res_\mathrm{exp}=\SI{240 \pm 1}{\micro\meter}$ close to the illuminating wavelength of $\lambda_\mathrm{THz}=\SI{200}{\micro\meter}$. This results in a number of $m_\mathrm{exp}=\SI{8 \pm 1}{}$ spatial modes per axis. For the theoretical value, we simulate the same knife-edge measurement and obtain a resolution limit of $res_\mathrm{sim}=\SI{174}{\micro\meter}$.
	
\begin{figure}
		\centering
		\makebox[\textwidth][c]{\includegraphics[width=\linewidth]{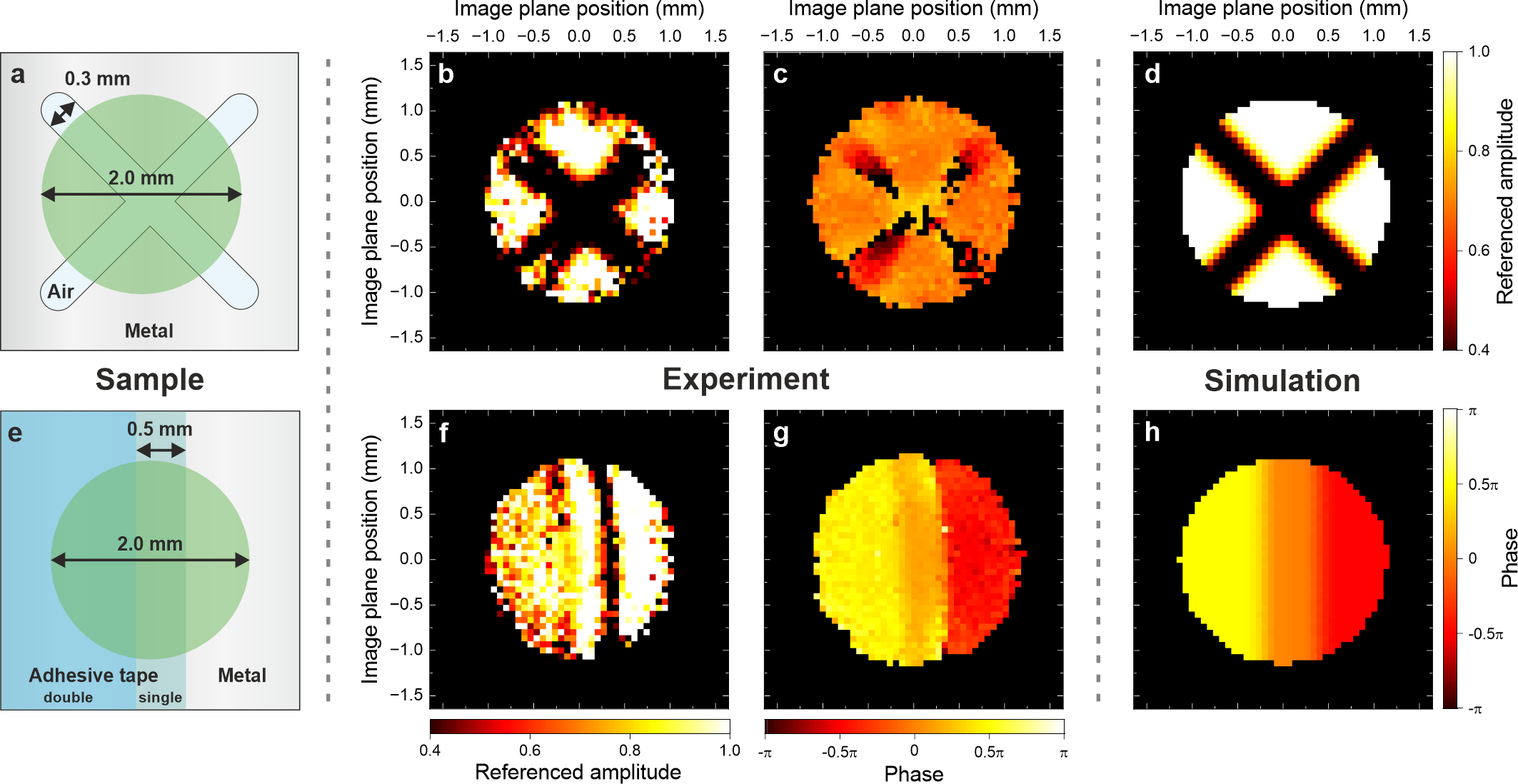}}
		\caption{Amplitude- and phase-based terahertz quantum imaging with undetected photons. (a) Schematic representation of the sample used for amplitude-based terahertz imaging (diagonal cross cutout) and the FoV of the incident terahertz beam (green). The cutout has a line width of \SI{0.3}{\milli\meter}. Measured amplitude- (b) and phase-based (c) image of the sample depicted in (a). (d) Simulation of the amplitude-based measurement. (e) Schematic representation of the sample used for phase-based terahertz imaging (vertical strips of double and single layer of adhesive tape) and the FoV of the incident terahertz beam (green). The width of the single-layer adhesive tape strip is about \SI{0.5}{\milli\meter}. Measured amplitude- (f) and phase-sensitive (g) image of the sample sketched in (e). (h) Simulation of the phase-based measurement. The amplitude-based images are related to the reference image shown in Fig.~\ref{fig:acquisition}d, for which a simple metal plate is used as a sample. The phase-based images indicate the phase shift resulting from the introduction of the sample relative to this reference image. For better visualization, the phase values in (g) and (h) are shifted by $-0.5\pi$.}
		\label{fig:ampandphase}
	\end{figure}

	In the following, various images are presented that demonstrate amplitude- and phase-sensitive quantum imaging in the terahertz frequency range. For this purpose, three different types of samples are manufactured, each emphasizing the accessibility to amplitude, phase, or spectral information of the terahertz radiation. For amplitude-based imaging, a metal plate modified with a cut-out area is used as sketched in Fig.~\ref{fig:ampandphase}a. The diagonal cross has a line width of \SI{0.3}{\milli\meter} while the whole structure exceeds the FoV, indicated by the green circle. The impinging terahertz photons are either reflected by the metal surface or transmitted through the cut-out areas. Interference occurs only in regions where the corresponding terahertz radiation is reflected and fed back into the nonlinear crystal. The amplitude-based image in Fig.~\ref{fig:ampandphase}b clearly shows the diagonal cross structure, although the sample features are on the order of the illumination wavelength.
	
	In addition to the amplitude-based evaluation, the phase shift introduced by the sample is determined by fitting a waveform model to the acquired data (Methods). The phase shift for the sample illustrated in Fig.~\ref{fig:ampandphase}a relative to the reference measurement is shown in Fig.~\ref{fig:ampandphase}c. In the areas of the metal plate, no observable phase shift is present. Although the sample structure is visible, it appears less distinct. We attribute this to the higher sensitivity of the phase evaluation to pixel binning compared to the amplitude-based evaluation.
	
	To assess the quality of our imaging system, we performed simulations of the detector images. In these simulations, we evaluated the features the sample intended to emphasize. The recorded amplitude-based image shows good agreement with the corresponding numerical simulation shown in Fig.~\ref{fig:ampandphase}d. The simulation shows a clearer image as no artificial noise is added, resulting in a lower noise level compared to the experimental results. The noise in the simulation is a numerical artifact due to the finite number of evaluated points. Further, the resolution of the simulated measurement is higher, as it assumes an ideal setup without misalignments.
	
	For phase-based imaging, two strips of adhesive tape were applied vertically to the metal plate, as sketched in Fig.~\ref{fig:ampandphase}e, dividing the area into three sections of different optical path lengths. The corresponding amplitude-based evaluation (see Fig.~\ref{fig:ampandphase}f) clearly shows the boundaries of the three areas due to the scattering of the terahertz radiation. Furthermore, the distinct extinction behavior in each region can be observed. The image in Fig.~\ref{fig:ampandphase}g shows the estimated phase shift introduced by the sample. For better visualization, the measured phases are shifted by -0.5$\pi$. All areas are clearly separated, and deviations of the phase shift estimated in one area are only apparent in regions of low interference amplitude (compare Fig.~\ref{fig:ampandphase}f). With a thickness of \SI{50}{\micro\meter} and a refractive index of about $n_\mathrm{T}$ = 1.5 at a frequency of \SI{1.5}{\tera\hertz}, a total phase shift of around 0.5$\pi$ for each layer of adhesive tape is expected. This shift is observed between the outer areas, one without and the other with a double layer of adhesive tape. However, the measured phase shift between the metal and a single layer of adhesive tape is about 2/3 $\pi$, while a phase shift of about $\pi$/3 is observed between the single and double layer of tape. We attribute the discrepancy in the center area to the influence of the surrounding areas and edge effects, which are significant due to its small width of \SI{0.5}{\micro\meter}. Nevertheless, this measurement shows a good agreement with the simulated image shown in Fig.~\ref{fig:ampandphase}h. In the simulation, the sample is assumed to be a pure phase object, for which no perturbing effects are considered, resulting in a clearer representation. As with the amplitude measurement, no artificial noise was simulated.
	
\begin{figure}
		\centering
		\includegraphics[width=80mm]{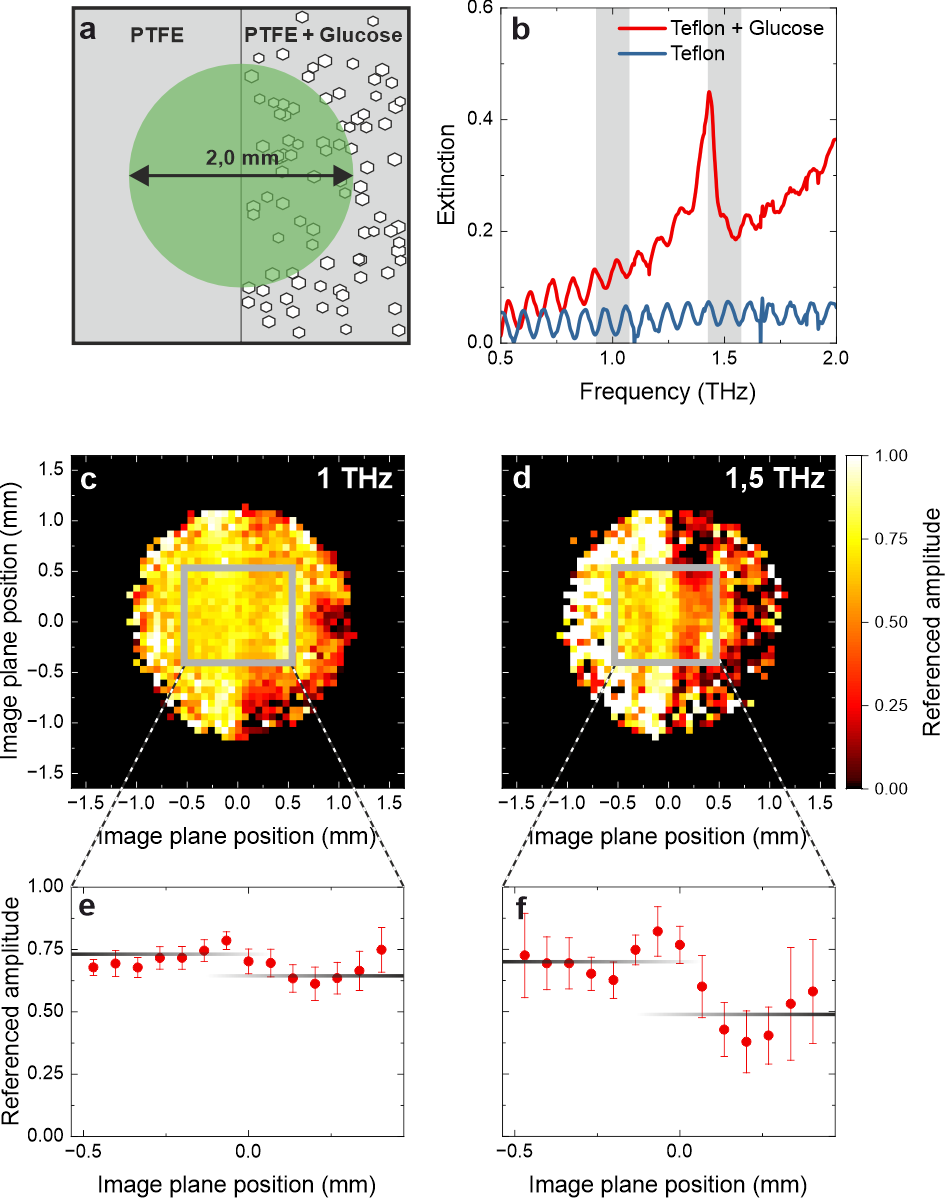}
		\caption{Terahertz quantum imaging based on spectral information. (a) Schematic of the used sample and the FoV of the incident terahertz beam (green). The sample is manufactured from two PTFE tablets, each with a thickness of about \SI{1}{\milli\meter}. To one tablet, glucose was added with a concentration of \SI{0.11}{\milli\gram\per\milli\meter\squared}. (b) Extinction measurement of the sample performed with a standard TDS system. The gray-shaded areas indicate the spectral regions of the measurements performed. Amplitude-based images of the sample taken at a center frequency of (c) \SI{1.0}{\tera\hertz} and (d) \SI{1.5}{\tera\hertz}. The gray box indicates the area evaluated quantitatively. For the quantitative evaluation of the measurement taken with a center frequency of (e) \SI{1.0}{\tera\hertz} and (f) \SI{1.5}{\tera\hertz}, the pixels are averaged vertically. The error bars represent the standard deviation.}
		\label{fig:absorption}
	\end{figure}

	To demonstrate imaging based on spectral information, we produced a sample with strong absorption in a specific area. Two polytetrafluoroethylene (PTFE) tablets are manufactured, with one modified by adding glucose. PTFE is an ideal carrier material for terahertz absorption experiments due to its high transmittance and low refractive index of 1.42 \cite{DAngelo.2014, Kutas.2020} in the terahertz frequency range. Glucose is chosen as an additive due to its well-known characteristic absorption feature at around \SI{1.45}{\tera\hertz} \cite{Upadhya.2003}. To ensure that only one-half of the sample contains the additive (see Fig.~\ref{fig:absorption}a), the tablets are bisected and merged together. Assuming a homogeneous distribution, the glucose concentration in the PTFE tablet is \SI{0.11}{\milli\gram\per\milli\meter\squared}.
	In Fig.~\ref{fig:absorption}b the extinction measured by a standard time-domain spectroscopy (TDS) system for both parts of the sample is shown. The observed oscillations originate from Fabry–Pérot interference within the tablet.
	
	In Fig.~\ref{fig:absorption}d, the amplitude-based image acquired with idler frequencies at \SI{1.5}{\tera\hertz} is shown. The measurement indicates significant attenuation for terahertz photons entering the right-hand side of the sample compared to the left. However, the associated frequency spectrum does not exhibit any distinct feature but rather a general reduction in amplitude due to the brief delay and the proximity to the water vapor absorption line at \SI{1.4}{\tera\hertz} (for a detailed explanation, see \cite{Kutas.2021}). To confirm that this behavior is primarily attributable to absorption introduced by the additive, a further measurement is carried out with idler frequencies around \SI{1.0}{\tera\hertz} for the same sample (see Fig.~\ref{fig:absorption}c). In this case, only a marginal difference is observed between the individual regions, as the extinction of PTFE in this frequency range increases only slightly with the addition of glucose (see Fig.~\ref{fig:absorption}b).
	
	The quantitative evaluation of the extinction experienced by the terahertz photons through the sample is limited to the area offering a high dynamic range. Assuming a homogeneous distribution of the additive, pixels in the vertical direction are expected to exhibit similar behavior. Therefore, only the vertical pixels of the marked area in the amplitude-based images are averaged, with error bars corresponding to the standard deviation of the distribution. Fig.~\ref{fig:absorption}f shows the result for the measurement taken at \SI{1.5}{\tera\hertz}, while Fig.~\ref{fig:absorption}e shows the result for the measurement at \SI{1.0}{\tera\hertz}.
	
	The difference between the two sides is apparent in both cases and is further highlighted by the horizontal lines, which result from transmission measurements carried out on the sample with a conventional TDS system. Since the optical systems of both setups are not comparable, only the transmission difference between the two parts of the sample is considered (vertical distance between the lines). The values are offset to align with the average measured value in the pure PTFE area (Methods).
	
	\section{Discussion}
	
	In this work, we have demonstrated quantum imaging with the largest energy spread between signal and idler photons reported to date, to the best of our knowledge. This paves the way for imaging in spectral ranges that are typically challenging to access. In the future, this technique could benefit from advancements in the rapidly evolving field of quantum sensing, which is still in its early stages, while also allowing for the integration of established classical imaging techniques. This dual potential offers numerous opportunities for future developments. Although current measurement times are not yet competitive to advanced TDS systems, our work could be an impetus for the further development of imaging methods in extreme spectral ranges, which often encounter similar challenges \cite{Kutas.2022}. Enhancement in resolution and measurement times could be achieved by using novel nonlinear materials like metamaterials \cite{Luo.2014} or by implementing acquisition schemes such as phase shift holography \cite{Topfer.2022, Haase.2023}. Moreover, employing external seed sources could significantly enhance the visibility and speed up the imaging process \cite{Cardoso.2018}.
	
	In conclusion, we have demonstrated amplitude- and phase-sensitive quantum imaging with undetected photons in the terahertz frequency range. The images are acquired using a quantum distillation scheme exhibiting a high resilience to noise. At a frequency of \SI{1.5}{\tera\hertz}, we achieve a resolution of \SI{240}{\micro\meter} with an associated FoV of up to \SI{2}{\milli\meter}. Our proof-of-principle images confirm that amplitude- and phase-based imaging as well as imaging based on spectral information can be realized.
	
	\newpage
	
	\section{Methods}
	
	\subsection{Experiment}
	
	\subsubsection{Biphoton source}
	The pump source is a linearly polarized, frequency-doubled, single longitudinal mode continuous-wave laser (Cobolt Flamenco\texttrademark). It has a center wavelength of \SI{659.58}{\nano\meter}, providing a narrow bandwidth of less than 1 MHz at an average output power of up to \SI{500}{\milli\watt}. The correlated biphoton pairs are generated in a \SI{1}{\milli\meter}-long periodically poled MgO:LiNbO$_3$ crystal. The nonlinear crystal provides two distinct areas with different poling periods of \SI{109.5}{\micro\meter} and \SI{72}{\micro\meter}, which generate photons with a center frequency of about \SI{1.0}{\tera\hertz} and \SI{1.5}{\tera\hertz}. Both areas are of a dimension of 4$\times$4 \si{\milli\meter} and separated by \SI{2}{\milli\meter} without periodic poling.
	
	\subsubsection{Parasitic signal radiation}
	
	The nonlinear generation of terahertz radiation using MgO:LiNbO$_3$ not only produces the required signal photons but also additional (parasitic) signal radiation. This can be observed especially when representing the generated signal radiation as frequency-angular spectrum (see \cite{Kitaeva.2011, Haase.2019}). We refer to this radiation as parasitic as it does not contribute to the interference we aim to observe in the experiment. The generation of this parasitic signal radiation has various causes. 
	First, the quasi phase-matching structure of the nonlinear crystal introduces an additional wave vector that must be considered in the phase matching conditions. This wave vector can be oriented in both the pump direction and its opposite. For terahertz radiation, both cases are phase-matched and lead to the generation of signal photons. The desired case for the experiment is when terahertz photons propagate in the direction of the pump. The opposite case, where terahertz photons are generated in the opposite direction, leads to parasitic signal radiation.
	Second, a nonlinear conversion of thermal terahertz radiation present at room temperature occurs. In this case, signal photons are generated in a down- as well as an upconversion process. Therefore, one part contributes to the considered wavelength in the experiment, but the main part leads to additional generated parasitic signal radiation. 
	Last, parasitic signal contributions are generated by Raman scattering in MgO:LiNbO$_3$.
	
	\subsubsection{Alignment of the nonlinear interferometer}
	
	The alignment of the experimental setup such that the biphotons generated in the first and second pass of the pump radiation are indistinguishable can not be carried out in real-time due to the low visibility of the interference. Therefore, the alignment is done in several steps. In the terahertz arm of the interferometer, a off-axis parabolic mirror (OAP) and a terahertz lens (made of polymethylpentene (TPX)) are used to build the imaging system. For the initial alignment of these components a scattering source is positioned at the crystal plane to align the OAP and the end mirror of the terahertz path so that the scattering source is imaged onto itself. To completely reflect the scattered light which is collimated after the OAP, the end mirror/sample used in the terahertz path has a dimension of $2\times2$ inches (see Fig.~\ref{fig:setup}). Afterward, the nonlinear crystal is positioned in the focal plane of the OAP. In addition to the pump radiation, the crystal is illuminated at this stage with external terahertz radiation using a photoconductive switch (PCS), employed exclusively for alignment and not for measurement. This PCS serves to seed the nonlinear conversion process in the crystal, where a part of the external terahertz radiation also propagates through following the terahertz path of the interferometer. This amplification leads to a higher number of generated signal photons in the crystal. The terahertz lens is then introduced into the beam path and positioned to maximize the signal strength measured on the camera. We then assume that the terahertz path is well-adjusted to observe interference. The signal arm of the interferometer is formed by the lenses f$_1$ and f$_2$, through which the signal and the pump pass. The alignment starts by adjusting the end mirror to reflect the pump radiation into a pinhole. Afterwards, the two lenses f$_2$ and f$_1$ are inserted into the beam path one after the other, whereby again, the beam is aligned with the previously positioned pinhole.
	
	Since the focal lengths of the lenses play a central role in the magnification of the imaging system, these are listed below. The lenses in the signal arm of the interferometer have a focal length of $f_1$=\SI{100}{\milli\meter} and $f_2$=\SI{30}{\milli\meter}, respectively. In the terahertz arm of the interferometer, the OAP provides a focal length of $f_\mathrm{{OAP}}$=\SI{50.8}{\milli\meter}, and the used lens has a focal length of $f_\mathrm{{THz}}$=\SI{65}{\milli\meter}. This results in a magnification of the terahertz spot imaged onto the detector of $M_\mathrm{THz}$=0.78.
	The total length of the optical system in the terahertz arm is slightly shorter (\SI{115.8}{\milli\meter}) than in the signal arm (\SI{130}{\milli\meter}). All lenses are aligned in a way that either the respective end mirror or the nonlinear crystal are positioned in the focal plane of the corresponding lens. The additional path length needed to balance the interferometer arms is introduced into the collimated part of the terahertz beam between the OAP and lens $f_\mathrm{{THz}}$.
	
	\subsubsection{Wavelength filtering}
	
	Volume Bragg gratings (BNF.660 from OptiGrate) are used as filters for the pump radiation. To suppress parasitic signal contributions, a spectral filter scheme is implemented. For spectral distribution of the signal, we use two highly efficient transmission gratings (PCG-1908-675-972 from Ibsen Photonics) with 1908 lines per millimeter and an efficiency greater than \SI{94}{\percent} at \SI{660}{\nano\meter}. 
	The spatial distribution of the signal, allows for the suppression of parasitic signal contributions that differ from the signal wavelength via a spatial filter (slit). Additonally, this scheme suppresses scattered pump radiation caused by the VBGs, remaining in the beam path due to the lenses used.
	
	\subsubsection{Detection}
	
	In the detection part of the nonlinear interferometer, several lenses are positioned as 4f imaging systems to image the signal onto the camera. The first 4f system is formed by two lenses with focal lengths $f_3$=\SI{300}{\milli\meter} and $f_4$=\SI{1000}{\milli\meter}. Subsequently, the lenses $f_5$ and $f_6$ form the next 4f system and have focal lengths of $f_5$=$f_6$=\SI{400}{\milli\meter}. The final 4f system reduces the size of the beam to image the signal onto the camera with focal lengths of $f_7$=\SI{400}{\milli\meter} and $f_8$=\SI{35}{\milli\meter}.This leads to a magnification of the signal spot from the crystal plane to the camera plane of $M_\mathrm{Vis} = 3.43$. For the magnification of the terahertz image taken in the sample plane to the camera, the resulting magnification is $M_\mathrm{image} = 2.67$.
	
	For signal acquisition, an uncooled sCMOS camera (Thorlabs Quantalux\textregistered~sCMOS Camera) with a specified quantum efficiency of \SI{55}{\percent} (at \SI{660}{\nano\meter}) is used. The pixel size is \SI{5.04}{\micro\meter} by \SI{5.04}{\micro\meter}, providing 2.1 megapixels with up to \SI{87}{\deci\bel} dynamic range. At \SI{20}{\celsius}, the labeled pixel dark count rate and readout noise are about 20 counts per second and 1 e$^{-}$, respectively. The measured background illumination of the presented results is about 150 counts per second and pixel (combining dark count rate, remaining stray, and ambient light).
	
	\subsubsection{Fourier transform of the waveform}
	After acquisition, the waveform for each single pixel is post-processed. Firstly, the waveform is offset by the mean value of the waveform to oscillate around zero. Secondly, the beginning and end of the recorded waveform are attenuated to zero. This has the advantage of reducing artifacts in the FFT that influence the measurement.
	
	\subsubsection{Phase evaluation}
	To determine the phase, the recorded waveform is fitted with the following function for every pixel. 
	\begin{equation}
		f(t) = A \cdot \sin(\nu \cdot t + \varphi) \cdot \exp\left(-\frac{1}{2} \cdot \frac{(t - t_\mathrm{c})^2}{w^2}\right) + y_0
	\end{equation}
	This function takes into account the sinusoidal interference pattern of the signal and its envelope resulting from the coherence length. $A$ represents the amplitude, $\nu$ the angular frequency, $\varphi$ the phase, and $y_0$ the offset of the sinusoidal interference signal. The shape of the envelope is determined by the center delay $t_\mathrm{c}$ and the width $w$. Apart from the amplitude and phase, all variables for the individual pixels are identical. Therefore, only these two are used in the fitting process.
	
	\subsubsection{Determination of the FoV and the resolution}
	
\begin{figure}
		\centering
		\includegraphics[width=88mm]{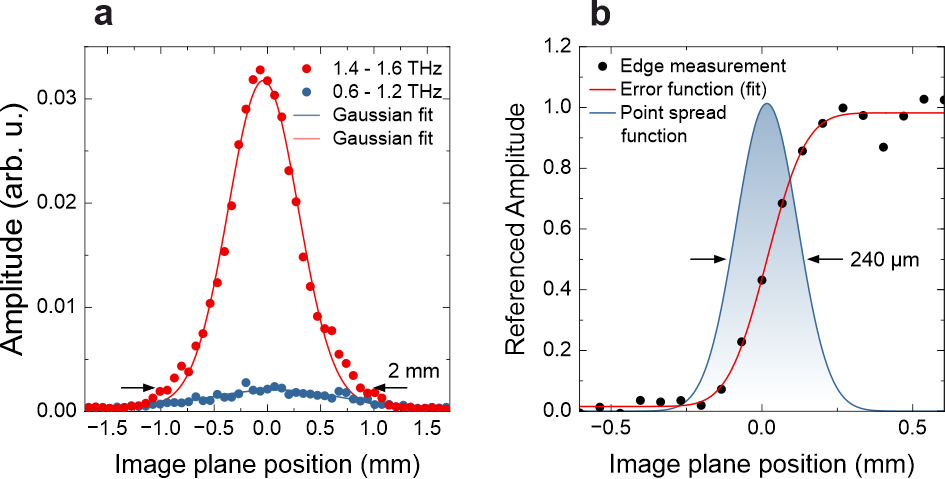}
		\caption{Characterization of the FoV and the resolution of the measurement system. (a) Cross-section of the reference image (red dots) and corresponding noise evaluation (blue dots). (b) The Measured edge response (dots) of the system is fitted with an error function (red line). The derivation of the error function leads to the corresponding point spread function (blue line). The resolution resulting from the full width at half maximum of the spread function is \SI{240}{\micro\meter}.}
		\label{fig:knive edge}
	\end{figure}

	For the presented imaging system, the FoV is limited by the noise level. To determine this noise level, the FFT of the waveform is evaluated for each pixel in a spectral region where no interference is observable (\SIrange{0.6}{1.2}{\tera\hertz}). In Fig.~\ref{fig:knive edge}a, a cross-section of the reference image (shown in Fig.~\ref{fig:acquisition}d) and the corresponding noise evaluation is shown. For terahertz images, only pixels with an amplitude of at least $1\times 10^{-3}$ above the noise level are taken into account. This results in an experimental $FoV_\mathrm{exp}$ of $\SI{2.0 \pm 0.2}{\milli\meter}$.
	
	To determine the resolution of the system, an edge-response measurement is carried out in the sample plane of the terahertz beam path. According to \cite{Kviatkovsky.2022}, the resolution of the system in position correlation is given by the point spread function. For better comparability, we will adhere to this definition. In Fig.~\ref{fig:knive edge}b, the corresponding edge response is presented. The measured values are fitted with an error function (red). The subsequent derivation, leads to the edge spread function (blue) of the system. As the resolution corresponds to the full width at half maximum of this spread function, we obtain a resolution of \SI{240 \pm 1}{\micro\meter}.
	
	\subsubsection{Terahertz time-domain spectroscopy measurements}
	
	For validation of the presented measurements based on spectral information, we performed measurements of the sample using a standard TDS system (bandwidth of about 4 THz, frequency resolution of 10 GHz). A detailed explanation of terahertz TDS spectroscopy can be found in \cite{Ellrich.2020}. To determine the extinction for both parts of the sample, we take a measurement of each half of the sample positioned in the focal plane between the emitter and the detector of the TDS system. As a reference, a measurement without the sample is used. This allows for the calculation of the frequency-dependent extinction as follows:
	
	\begin{equation}
		K(f) = 2 \log_{10}\left(\frac{A_0(f)}{A(f)}\right).
	\end{equation}
	
	$A_0(f)$ is the frequency-dependent amplitude of the reference taken without the sample and $A(f)$ the frequency-dependent amplitude of the measurement taken with the sample. In Fig.~\ref{fig:absorption}b the extinction measured for each part of the sample is shown in dependence of the terahertz frequency.
	
	When comparing the measurement of the TDS system with our imaging system, the following differences must be considered. First, the TDS system measures the electric field of the terahertz radiation, while our presented imaging system measures its intensity. Second, measurements with the TDS system are carried out in a transmission configuration, and the optical elements (apertures and focal lengths) used in both measurement systems differ. Third, the TDS system provides only a single-pixel acquisition, averaging over a larger area of the sample. Therefore, measurements taken with both systems are not directly comparable.
	
	To establish comparability, we evaluate the referenced amplitude of the TDS system by limiting it to the respective spectral range of the source used. Since the TDS system measures the amplitude of the electric field, these values must be squared to be comparable to the values from our imaging system. As the optical elements of the two systems differ, we focus only on the deviation of the values measured with the TDS system for both parts of the sample. For better comparability, we shift the values acquired with the TDS system for the pure PTFE sample part to match the average of the values measured with our imaging system for the same part of the sample. Therefore, these values (horizontal lines in Fig. \ref{fig:absorption}e and Fig. \ref{fig:absorption}f) provide an estimation of the expected reference amplitude measured with our system.
	
	\subsection{Theory and Simulation}
	
	The FoV of a quantum imaging setup in position correlation configuration depends on the beam radius with which the crystal is initially illuminated by the pump radiation and on the magnification of the lens system that illuminates the sample with the terahertz radiation. The corresponding values determined in the experiment are $\omega_\mathrm{p}=\SI{0.89}{\milli\meter}$ for the pump beam waist illuminating the crystal and $M_\mathrm{THz}=0.78$ for the magnification of the imaging system. According to \cite{Kviatkovsky.2022}, the FoV  depends on the beam waist of the pump beam the nonlinear crystal is illuminated with according to
	
	\begin{equation}
		\mathrm{FoV}=\sqrt{2\ln{2}}\,\frac{\omega_\mathrm{p}}{M_\mathrm{THz}}.
	\end{equation}
	
	With a magnification $M_\mathrm{THz}=0.78$ and a beam waist of $\omega_\mathrm{p}=\SI{0.885}{\milli\meter}$, this results in a calculated $FoV_{\mathrm{theo}}=\SI{1.3}{\milli\meter}$. However, as described in \cite{Kviatkovsky.2022}, evaluating the pixel-by-pixel visibility not only allows us to measure phase and amplitude information, but also results in a larger effective FoV, as the visibility distribution is flatter. If the image is additionally related to a reference measurement, even smaller changes can be made visible. This is particularly advantageous for low visibilities and makes the illuminated area of around \SI{2.0}{\milli\meter} almost completely usable as the FoV.
	
	For a more accurate description of the measurement system, which also reflects the pixel-wise approach, we use a numerical simulation method for quantum sensing systems \cite{Riexinger.2023b, Riexinger.2023}. The method allows for the simulation of detector images for both phase and transmission objects. We adapt this method to position-correlated imaging by using position instead of momentum states. A transition amplitude for photon pairs with transverse positions $\rho_{\mathrm{Vis}}$ and $\rho_{\mathrm{THz}}$ can be derived by making the paraxial approximation and assuming perfect transversal momentum correlation \cite{Viswanathan.2021}. The resulting transition probability is given as
	\begin{align}
		P(\rho_{\mathrm{Vis}}, \rho_{\mathrm{THz}}) & = \frac{8}{\pi L \omega_{\mathrm{p}}^2 (\lambda_{\mathrm{Vis}} + \lambda_{\mathrm{THz}})} \exp\left(-\frac{4\pi}{L(\lambda_{\mathrm{Vis}} + \lambda_{\mathrm{THz}})} \left|\rho_{\mathrm{Vis}} - \rho_{\mathrm{THz}}\right|^2\right) \nonumber\\ &\times \exp\left(-\frac{2}{\omega_{\mathrm{p}}(\lambda_{\mathrm{Vis}} + \lambda_{\mathrm{THz}})^2} \left|\lambda_{\mathrm{THz}}\rho_{\mathrm{Vis}} - \lambda_{\mathrm{Vis}}\rho_{\mathrm{THz}}\right|^2\right),
	\end{align}
	with the crystal length $L$, and the wavelengths of the detected ($\lambda_{\mathrm{Vis}}$) and undetected ($\lambda_{\mathrm{THz}}$) photons. 
	For an ideal imaging system we then obtain the count rate at a detector pixel with center $\rho_{\mathrm{D}}$ as
	\begin{align}
		\label{eq:coun_rate}
		R(\rho_{\mathrm{D}}) \propto \int \mathrm{d} \rho_{\mathrm{THz}} P(\frac{\rho_{\mathrm{D}}}{M_{\mathrm{Vis}}}, \rho_{\mathrm{THz}}) \left[ 1 + |T(\rho_{\mathrm{THz}} M_{\mathrm{THz}})| \cos(\phi(\rho_{\mathrm{THz}} M_{\mathrm{THz}}) + \phi_{\mathrm{exp}})\right],
	\end{align}
	where $T$ and $\phi$ are the transmissivity and the phase introduced by the object, $\phi_{\mathrm{exp}}$ is the relative phase introduced by the variable delay, and we made the assumption that the propagation through the optical system can be described by its magnification in the object ($M_{\mathrm{THz}}$) and detector ($M_{\mathrm{Vis}}$) plane. 
	
	To evaluate the integral in equation~(\ref{eq:coun_rate}) we use a quasi-Monte Carlo integration scheme, exploiting the fact that the transition probability can be separated into to Gaussian terms. The propagation of the biphoton states through the setup to the object and detector plane is done with ray optics. The changes the object introduces to the biphoton states are represented by the transmissivity and phase terms in equation~(\ref{eq:coun_rate}).
	
	To simulate the amplitude and phase measurements, as well as the characterizing quantities, we reproduce the same steps as those preformed in the experimental setup.
	
	From the simulations, we obtain the resolution of the system analogously to the experimental method as $res_\mathrm{sim}=\SI{174}{\micro\meter}$, which provides a theoretical limit. However, the simulations do not include the finite aperture of the OAP, which is completely illuminated by the generated terahertz radiation. The loss of idler radiation, which is not reflected by the OAP, leads to a lower resolution limit. To reflect this, one can follow the analytical approach of  \cite{Kviatkovsky.2022}, which considers the apertures of the system:
	
	\begin{equation}
		res=0.51 \frac{\lambda_\mathrm{THz}}{\mathrm{NA}_\mathrm{lim}}.
	\end{equation}
	
	Where $\mathrm{NA}_\mathrm{lim}$ indicates the limiting numerical aperture in the experiment and $\lambda_\mathrm{THz}$ the terahertz wavelength. Using the numerical aperture of the OAP yields a theoretical resolution of $res_{\mathrm{theo}} \approx \SI{230}{\micro\meter}$ for our system at a frequency of \SI{1.5}{\tera\hertz}.

	\subsection*{Acknowledgements}
	
	We thank M. Gilaberte Basset and R. Sondenheimer for fruitful discussions. We thank D. Kharik for the fabrication of the amplitude-based sample. This project was funded by the Fraunhofer-Gesellschaft within the Fraunhofer Lighthouse Project Quantum Methods for Advanced Imaging Solutions (QUILT). 
	
%
%
%
%
%
%
%
%
%



\end{document}